\documentclass[12pt]{article}
\usepackage{amssymb}
\usepackage{amsmath}

\input{tcilatex}
\begin{document}

\noindent \textbf{A Link Mixture Model for Spatio-temporal Infection Data,
with Applications to the COVID\ Epidemic}\newline
\newline
P.Congdon, School of Geography, Queen Mary University of London, London E1
4NS. p.congdon@qmul.ac.uk\newline
\newline
\textbf{Abstract}. Spatio-temporal models for infection counts generally
follow themes of the broader disease mapping literature, but may need to
address specific features of spatio-temporal infection data including
considerable time fluctuations (with epidemic phases) and spatial diffusion.
Low order autoregression is a feature of several recent spatio-temporal
studies of infection data, possibly with lags on both within area infections
and on infections in adjacent areas. Many epidemic time series show a period
of relatively stable infection levels (possibly characterized as
endemicity), followed by a sudden sharp phase of increasing infection
levels. After the epidemic peak there is a period of descending rates and
return to stability. Hence one may seek to adapt the autoregressive scheme
to these pronounced fluctuations, with temporary departures from
stationarity, but returning to stationarity as rates descend and infections
resume endemic levels. We consider a mixture link model for infection counts
that allows adaptivity to both explosive phases and static endemicity. Two
case study applications involve COVID\ area-time data, one for 32 London
boroughs since the start of the COVID\ epidemic, the other focusing on the
epidemic phase in 144 area of South East England associated with the Delta
variant. \newline
\newline
\textbf{Keywords} Autoregressive. COVID. Endemic. Epidemic. Link mixture.
Spatio-temporal\newline
\newline
\textbf{Introduction\newline
}\newline
There have been a considerable number of spatio-temporal studies of disease
patterns, often adopting a Bayesian perspective (e.g. Lagazio et al, 2021).
Spatio-temporal models for infection counts (e.g. Lowe et al, 2021; Coly et
al, 2021; Jalilian and Mateu, 2021; Lawson and Song, 2010) are a particular
sub-theme. These may incorporate the themes of the broader disease mapping
literature, such as the gains through borrowing strength, and the need to
reflect spatial correlation in disease; for example, see Andrews et al
(2021) on spatial clustering in COVID rates.\newline
\newline
However, spatio-temporal infection data raises particular new issues. These
include considerable time fluctuations (with epidemic phases) in many
infectious diseases; spatial diffusion or spillover (e.g. Dalvi and Braga,
2019) related to behaviours such as commuting (Mitze and Kosfeld, 2021), and
the occurrence of hotspots for disease propagation early in infectious
outbreaks (Dowdy et al, 2012). It is also especially useful in policy terms
to be able to extrapolate the infectious disease evolution beyond the
observation span, as illustrated in some studies of the COVID epidemic
(Vahedi et al, 2021; Rui et al, 2021; Watson et al, 2017; Giuliani et al,
2020).\newline
\newline
Low order autoregression is a feature of several recent spatio-temporal
studies of infection data. For example, Paul and Held (2011) and Shand et al
(2018) adopt first order autoregression (AR1) models, where the
autoregressive coefficients on counts or rates in the previous period are
spatially correlated. The model of Paul and Held (2011) includes a spatial
lag on infection counts in adjacent areas that allows for the spatial
spillover effect; related approaches are considered by Martines et al (2021)
and Griffith and Li (2021). In infections spread by human contact it is
implausible that higher counts or rates in one period generate smaller
infection levels in the next period, and so a positive constraint on the AR1
coefficient may be adopted. Stationarity may also be assumed (e.g. Shand et
al, 2018), but flexibility to epidemic fluctuations may be gained by
allowing non-stationarity. \newline
\newline
A particular feature of epidemic time series is that a period of relatively
stable infection levels (possibly seen as an endemic phase) is followed by a
sudden sharp phase of increasing infection levels. After the epidemic peak
there is a period of descending rates and return to stability. Hence one may
seek to allow the autoregressive scheme to adapt to these pronounced
fluctuations, with temporary departures from stationarity, but returning to
stationarity as rates descend and infections resume endemic levels. For
example, assuming an identity link in count regression, AR1 coefficients
exceeeding 1 may better represent and reproduce sharply increasing infection
levels during an explosive phase. In this paper we consider a mixture link
model for infection counts that allows such adaptivity to both explosive
phases and static endemicity. \newline
\newline
Two case study applications involve area-time data for COVID-19 infection
counts. The first case study considers the 32 London boroughs which
exemplify the sudden growth in COVID infections due to the Omicron variant
at the end of 2021. The link mixture approach is applied to data from the
start of the epidemic in March 2020 through to early 2022 for the London
boroughs, and shows better fit than simpler options and improved short term
forecasts. The second case study considers data for the South East of
England (144 areas), and considers especially the weeks up to and including
the peak of infections due to the Delta variant at the end of 2020. \newline
\newline
\textbf{Methods}\newline
\newline
\textbf{Autoregression for Infection Counts}\newline
\newline
Consider area-time infection count data $y_{it}$ for areas $i=1,...,N$ and
times $1,...,T$, and assume these are negative binomial (NB), $y_{it}$ $%
\thicksim NB(\mu _{it},\Psi ).$ The NB parameterisation is

$p(y|\mu ,\Omega )=\frac{(y+\Psi -1)!}{y!(\Psi -1)!}\left( \frac{\mu }{\mu
+\Psi }\right) ^{y}\left( \frac{\Psi }{\mu +\Psi }\right) ^{\Psi }$.\newline
\newline
Assume an AR1 model on previous infection counts in the same area.
Additionally stable effects of predictors $X_{i},$ and constant unobserved
area effects $u_{i}$ (random or fixed), may be represented by a term $\eta
_{i}=X_{i}\beta +u_{i}.$ Then for a basic model, and conditioning on the
first period's data, one may adopt an identity link

$\mu _{it}=\rho _{i}y_{i,t-1}+\exp (\eta _{i}),\qquad \qquad (1)$\newline
providing positivity in $\rho _{i}$ is ensured. The $\rho _{i}$ may well be
spatially correlated, and typically taken as random effects.\newline
\newline
One may add lags to infection counts in spatially close areas to reflect
infection spillover from neighbouring areas - due, for example, to social
interactions between residents in different areas, or to cross boundary
commuting (Mitze and Kosfeld, 2021). Let $h_{ij}$ measure spatial
interactions between areas $i$ and $j$, and $w_{ij}=h_{ij}/\underset{j}{\sum 
}h_{ij}$ be row standardised spatial weights, with $\sum_{j}w_{ij}=1$. Then
spatial spillover, also with lag 1, can be represented (e.g. Paul and Held,
2011) by adding a spatially averaged term $\lambda
_{i}\sum_{j}w_{ij}y_{j,t-1}$ to the above basic model. Then one has

$\mu _{it}=\rho _{i}y_{i,t-1}+\lambda _{i}\sum_{j}w_{ij}y_{j,t-1}+\exp (\eta
_{i}),\qquad \qquad (2)$\newline
providing positivity in $\rho _{i}$ and $\lambda _{i}$ is ensured.\newline
\newline
Assuming that $\rho _{i}$ and $\lambda _{i}$ are positive is justified in
epidemiological terms, since - for infections spread by human contact or
interaction - higher current totals of infectees, $y_{i,t-1},$ and $%
\sum_{j}w_{ij}y_{j,t-1},$ are expected to cause higher future infections. A
negative effect of existing infection levels on future infections is
implausble. \newline
\newline
One then requires an appropriate link function relating $\rho _{i}$ and $%
\lambda _{i}$ to relevant parameters. For example, assume spatially
correlated conditional autoregressive random effects $f_{1i}$ and $f_{2i}$
(Besag et al, 1991) involved in predicting $\rho _{i}$ and $\lambda _{i}$,
and assume these are zero-centred. Paul and Held (2011, p. 1121) adopt a log
link by default, so that with intercept terms $\alpha _{1}$ and $\alpha _{2}$%
, one has

$log(\rho _{i})=\alpha _{1}+f_{1i},\qquad \qquad (3)$

$log(\lambda _{i})=\alpha _{2}+f_{2i}.\qquad $\newline
A log link allows for explosive effects ($\rho _{i}$ and/or $\lambda _{i}$
exceeding 1), but does not necessarily select explosive behaviour. Whether
it does will depend on the estimated values of the intercepts and spatial
effects. If most of the epidemic series consists of stable infection levels
(endemicity) then the estimated constants may tend to favour\ $\rho _{i}$
and $\lambda _{i}$ below 1.\newline
\newline
For some infectious diseases, with endemic recurrence now predominant, such
as HIV in developed countries, a stationary autoregressive effect may be
appropriate as a default. See, for example, the analysis of Shand et al
(2018) who consider time variations in HIV\ in US\ counties. For an AR1
model on lagged infections, this implies a logit link so that $\rho _{i}$
and $\lambda _{i}$ are constrained between 0 and 1. Thus, with the same
overall model $(2)$ for counts, and spatial effects $f_{3i}$ and $f_{4i},$
one has

$logit(\rho _{i})=\kappa _{1}+f_{3i},\qquad \qquad (4)$

$logit(\lambda _{i})=\kappa _{2}+f_{4i},$\newline
with expit equivalents

$\rho _{i}=1/(1+\exp (-\kappa _{1}-f_{3i}))$

$\lambda _{i}=1/(1+\exp (-\kappa _{2}-f_{4i}))$\newline
\newline
In fact, if positivity of $\rho _{i}$ and $\lambda _{i}$ cannot be assumed
on substantive grounds(e.g. for infections such as malaria not spread by
human interaction), then a mapping to the interval [-1,1] for $\rho _{i}$
and $\lambda _{i}$ can be obtained. If $h$ is the argument of the expit
transform, then the mapping $2expit(h)-1$ is to the interval [-1,1].\newline
\newline
\textbf{A\ Link Mixture Mechanism\newline
}\newline
For infections such as COVID, both mechanisms may be relevant. A logit link
may be relevant when infections are at a low and/or stable level, whereas a
log link, allowing $\rho _{i}>1$ and $\lambda _{i}>1,$ may be more flexible
in periods with explosive growth in infections (e.g. due to a new virus or
new variants of that virus). An example is the rapid increase in COVID
infections linked to the emergence of the Omicron variant, as considered in
the first case study.\newline
\newline
Here we consider a mixture model facilitating time-variation in which link
is predominant, so reflecting the current infection phase. Other forms of
mixing between links have been considered, or extra parameters introduced
into modelling links. For example, Lang (1999) considers a mixture of the
canonical symmetric logistic link and one or more asymmetric forms in
modelling ordinal and binary outcomes, while Czado and Raftery (2006)
consider right and/or left tail modifications to standard links.\newline
\newline
Here we consider a situation not researched before (as far as the authors
know), namely choosing between log and logit links. Thus, for weights $%
\omega _{t}$ between 0 and 1, it is proposed that

$\rho _{it}=\omega _{t}\exp (\alpha _{1}+g_{1i})+(1-\omega _{t})$ $\frac{%
\exp (\kappa _{1}+g_{1i})}{1+\exp (\kappa _{1}+g_{1i})},\qquad \qquad (5)$

$\lambda _{it}=\omega _{t}\exp (\alpha _{2}+g_{2i})+(1-\omega _{t})$ $\frac{%
\exp (\kappa _{2}+g_{2i})}{1+\exp (\kappa _{2}+g_{2i})},$\newline
where $\rho _{it}$ and $\lambda _{it}$ now vary by area and time, and $%
g_{1i} $ and $g_{2i}$ are spatially correlated conditional autoregressive
random effects. The $\omega _{t}$ are in effect measuring stability or
instability in infection rates, and so are taken as common to both own area
and the neighbouring area lags, $\rho _{it}$ and $\lambda _{it}$
respectively. For $\omega _{t}$ high and approaching 1, infections are
typically rapidly increasing, whereas for low $\omega _{t},$ stable
endemicity is indicated. Low $\omega _{t}$ may also be better for
characterizing the descent phase after epidemic peaks.\newline
\newline
If the stationary alternative to explosive growth involves the constraints $%
\rho _{it}$ $\in $ $[-1,1],$ and $\lambda _{it}\in $ $[-1,1],$ rather than $%
\rho _{it}$ $\in $ $[0,1],$ and $\lambda _{it}\in $ $[0,1],$ this can be
achieved by the mapping

$\rho _{it}=\omega _{t}\exp (\alpha _{1}+g_{1i})+(1-\omega _{t})$ $\left( 2%
\frac{\exp (\kappa _{1}+g_{1i})}{1+\exp (\kappa _{1}+g_{1i})}-1\right)
,\qquad \qquad $

$\lambda _{it}=\omega _{t}\exp (\alpha _{2}+g_{2i})+(1-\omega _{t})$ $\left(
2\frac{\exp (\kappa _{2}+g_{2i})}{1+\exp (\kappa _{2}+g_{2i})}-1\right) .$%
\newline
\newline
There is no reason why spatial patterning in autocorrelation should be the
same in epidemic or endemic phases, so a possible variation on the preceding
model allows for differing spatial effects between phases, namely

$\rho _{it}=\omega _{t}\exp (\alpha _{1}+g_{1i})+(1-\omega _{t})$ $\frac{%
\exp (\kappa _{1}+g_{3i})}{1+\exp (\kappa _{1}+g_{3i})},\qquad \qquad (6)$

$\lambda _{it}=\omega _{t}\exp (\alpha _{2}+g_{2i})+(1-\omega _{t})$ $\frac{%
\exp (\kappa _{2}+g_{4i})}{1+\exp (\kappa _{2}+g_{4i})}.$\newline
Under both $(5)$ and $(6)$, focussing on area variations in $\rho _{it}$ and 
$\lambda _{it}$ during periods with explosive growth will indicate which
areas have been more subject to such growth. The summary coefficients $%
\overline{\rho }_{t}$ and $\overline{\lambda }_{t}$, obtained by averaging $%
\rho _{it}$ and $\lambda _{it}$ over areas, give an overall impression of
infection growth or endemic phases. \newline
\newline
The $\rho _{it}$ and $\lambda _{it}$ can also be compared to the threshold
of 1 to give an probability indication of explosive growth in different
areas. Thus define indicators

$r_{it}^{x}=I(\rho _{it}>1),$

$l_{it}^{x}=I(\lambda _{it}>1),\newline
$from which area-time exceedance probabilities can be estimated. Also the
sums $R_{t}^{x}=\sum_{i}r_{it}^{x}$ and $L_{t}^{x}=\sum_{i}l_{it}^{x}$ show
total areas with explosive infection growth in each period.\newline
\newline
The $\omega _{t}$ in $(5)$ and $(6)$ are modelled as time-specific beta
variables

$\omega _{t}\sim Beta(q_{1t},q_{2t}),$\newline
where $q_{1t}$ and $q_{2t}$ are positive parameters. Relevant covariates if
available (e.g. the proportions of infections due to a new variant) may be
used in predicting the $\omega _{t}$ via beta regression. Intervention
variables may also be included in this regression.\newline
\newline
The models in $(2)$ may be extended to include time and area-time varying
effects, such as seasonal effects, or unobserved area-time random effects $%
\delta _{it}$. These represent local trends not fully captured by
autoregressive effects on lagged infection levels. Thus for representations $%
(3)$ and $(4)$, one has

$\mu _{it}=\rho _{i}y_{i,t-1}+\lambda _{i}\sum_{j}w_{ij}y_{j,t-1}+\exp (\eta
_{i}+\delta _{it}),\qquad \qquad (8)$\newline
while for representations $(5)$ and $(6)$, one has

$\mu _{it}=\rho _{it}y_{i,t-1}+\lambda _{it}\sum_{j}w_{ij}y_{j,t-1}+\exp
(\eta _{i}+\delta _{it}).\qquad \qquad (9)$\newline
\newline
\textbf{Model Specification\newline
}\newline
The forms $(8)$ and $(9)$ are adopted in the case studies below. The spatial
effects $(f_{1i},f_{2i},f_{3i},f_{4i})$ and $(g_{1i},g_{2i},g_{3i},g_{4i})$
involved in defining the autoregression coefficients are taken to follow the
conditional autoregressive (CAR) scheme of Besag et al (1991). It is assumed
that

$\eta _{i}=X_{i}\beta +u_{i}$,\newline
where $u_{i}$ are also mean centred CAR\ spatial effects$.$ It is assumed
that the area-time effects $\delta _{it}$ follow a first order random walk $%
\delta _{it}\sim N(\delta _{i,t-1},\sigma _{\delta }^{2}),$ with initial
conditions $\delta _{i1}$ taken as fixed effects, $\delta _{i1}\thicksim
N(0,1)$. For identification an intercept is omitted from $X_{i}\beta ,$ and
covariates are centred. A\ single covariate is used in both case studies,
the mid-2020 population estimates, divided by 100 thousand.\newline
\newline
Gamma priors with shape one, and rate $0.01,$ are adopted on inverse
variance parameters, the parameters $\{q_{1t},q_{2t}\}$, and on the negative
binomial overdispersion parameter $\Omega $. Normal $\mathcal{N}(0,100)$
priors are assumed on fixed effects $\{\alpha _{1},\alpha _{2},\kappa
_{1},\kappa _{2},\beta _{1}\}.$\newline
\newline
We consider one step ahead predictions. The predictive means are taken as

$\mu _{i,T+1}=\rho _{iT}y_{i,T}+\lambda _{iT}\sum_{j}w_{ij}y_{j,T}+\exp
(\eta _{i}+\delta _{i,T+1}),$ \newline
and include the updated value $\delta _{i,T+1}\thicksim N(\delta
_{iT},\sigma _{\delta }^{2})$.\newline
\newline
\textbf{Analysis and Estimation\newline
}\newline
We apply the link mixture models specified in equations $(5)$ and $(6)$, and
mean as in $(9),$ these constituting models 3 and 4 respectively. Two
simpler options are the log link as in $(3)$, constituting model 1, and the
other (as model 2) is the logit link, as in $(4)$. \ Models 1 to 4 are
denoted M1, M2, M3 and M4 respectively. Bayesian estimation is adopted, and
implemented via the BUGS program (Lunn et al, 2009). Two chains of 20,000
iterations are taken, with inferences from the last 10,000, and convergence
checks as in Brooks and Gelman (1998). \newline
\newline
Fit is measured by the widely applicable information criterion (WAIC)
(Watanabe, 2013). Performance of predictions $P(y_{rep,it}|y_{it})$ =$\dint
P\left( y_{rep,it}|\theta \right) P(y_{it}|\theta )d\theta $ (where $\theta $
denotes all parameters) is measured by the Dawid-Sebastiani score (DSS)\ and
by the ranked probability score (RPS) (Czado et al, 2009). Let $Y_{t}$
denote region wide totals at period $t$ (i.e. total infections for all areas
combined). Assume the models are fitted to $T$ time periods, with period $%
T+1 $ as hold out. One step ahead predictions to $T+1$ are assessed by
whether these predictions include actual infection counts at $T+1$, and by
the RPS for one step ahead predictions.\newline
\newline
\textbf{Case Study 1:\ London Boroughs, 32 areas, 96 weeks}\newline
The data for the first study consist of weekly totals of new COVID\ cases in
the 32 boroughs of London. The time span considered starts at the week
ending Sunday, 8 March 2020 (constituting week 1). The final observation
(week 96) is that ending Sunday 2nd January 2022. Figures 1A and 1B show the
trajectory of total cases, in two successive sub-periods. The upturn due to
Omicron is apparent in the last few weeks of the series. The peak infections
were at week 94 (with 169322 cases, compared to 65771 in the previous week),
after which a downturn started. At the peak of the London Omicron wave, the
UK\ Office of National Statistics estimated that around 8.8\% of Londoners
had COVID-19 (ONS, 2022, Table 1e). An earlier upturn due to the Delta
variant produced a peak infection count of 93798 for the week ending
03/01/2021 (week 44 of the series), with a lesser upturn peaking at week 72.
The peak in infections early on in the epidemic (peaking at week 5) is
dwarfed by later upturns.\newline
\newline
We take weeks 1-95 as the observed data, with week 96 as hold out. There
were 155181 cases in that week, as infection levels due to Omicron started
to tail off from the peak at week 94. Table 1 compares the four models in
terms of fit to the data and prediction accuracy within the observed span.
Table 1 also compares their out of sample predictions to week 96. \newline
\newline
Regarding fit to the observed data, the WAIC, RPS and DSS criteria are all
lower for the link mixture models than for the log and logit models
(equations 3 and 4 respectively). Figure 2 shows the posterior mean RPS\
values for models M1-M4 disaggregated by week, with worse predictions under
models 1 and 2 showing especially in epidemic phases (M1 and M2 are the red
and blue lines in Figure 2). \newline
\newline
Models 3 and 4 also have greater accuracy in one step ahead prediction, in
terms of the coverage of the 95\% credible interval (crI) for $Y_{rep,T+1}$
of the actual value, and the RPS\ for week $T+1$. The 95\% crI under model 4
is (154018, 175048) including the true value, and the one-step ahead RPS is
42923. The 95\% crI under model 3 also includes the true value. By contrast,
the one step ahead RPS posterior means for M1 and M2 are 96920 and 95520.
These models over-predict $Y_{T+1}$, and the 95\% credible intervals for $%
Y_{rep,T+1}$ under M1 and M2 do not include the actual value.\newline
\newline
The posterior mean $\rho _{i\newline
}$ under M1 (which are time constant) vary from 0.13 to 0.25, while the mean 
$\lambda _{i}$ vary from 0.19 to 0.72. The higher values show which boroughs
tended to have higher growth in cases over one or more epidemic upturns. By
contrast, under models 3 and 4, $\rho _{it}$ and $\lambda _{it}$ parameters
often exceed 1 in weeks with high growth in cases, and one may identify the
upturn weeks where particular areas have higher growth. Table 2 accordingly
shows the 20 weeks with the highest values of $R_{t}^{x}$ under M4$.$ In a
few weeks (such as weeks 2 and 94), all 32 boroughs have nonstationary
growth, but Table2 shows that such growth is concentrated in a relatively
few weeks in the observation span. \newline
\newline
Figures 3A, 3B, and 3C plot out the posterior mean $\omega _{t}$ under M4
for weeks 50-95, the observed relative growth ratios $Y_{t}/Y_{t-1}$%
\thinspace\ and the averages $\overline{\rho }_{t}$ of the $\rho _{it}.$
Plots including the full observation span are unduly dominated by the
extremely high relative growth rate $Y_{2}/Y_{1}$, namely 8.3, as weekly
infections in London rose sharply when the COVID epidemic started in March
2020. In fact for weeks 50-95, the $\omega _{t}$ and $\overline{\rho }_{t}$
correlate highly (over 0.99) with the ratios $Y_{t}/Y_{t-1}.$ For 6 of these
45 weeks, the London wide $\overline{\rho }_{t}$ under M4 have posterior
mean exceeding 1, with the highest $\overline{\rho }_{t}$ being 1.81 for
week 94. These results confirm the utility of the link mixture mechanism in
reproducing actual infection count fluctuations. \newline
\newline
\textbf{Case Study 2: South East England, 144 areas, 20 weeks.\newline
}The data for this study relate to the broader South East of England,
encompassing 144 local authority areas in three standard regions (London,
East, and South East). The time span consists of 21 weeks from the week
ending the 9th August 2020 through to the week ending 27th December 2020.
This period includes a peak in cases related especially to the Delta
variant, namely week 21 with 210099 cases. Figure 4 shows a lesser peak at
week 14. Relative increases in weeks 7 and 8 are also high (over 60\%), but
case numbers remained low as compared to later weeks. We consider
observations for the first 20 weeks, with week 21 held out from estimation.
We compare the four models in terms of their fit to the observed data (weeks
1-20), and one step ahead predictions to week 21 when cases peaked.\bigskip 
\newline
Table 3 shows, as for the London study, that models 3 and 4 provide better
fit and predictions to the observed data. Table 4 shows the RPS\ by week for
the four models. Models M1 and M2 have worse predictive fit in weeks with
rapid shifts in case numbers (large increases or falls, as in weeks 19 and
15). As to tracking extreme increases associated with the Delta variant,
Figures 5A and 5B plot out the M4 posterior means by period of the
statistics $R_{t}^{x}$ and $L_{t}^{x},$ the number of areas with slopes $%
\rho _{it}$ or $\lambda _{it}$ exceeding 1. These both peak in week 19, at
44.5 and 40.8 respectively (out of a total of 144 areas), implying that
sharp growth in cases is from both local transmission and broader geographic
diffusion. Comparison with Figure 5C shows how these statistics closely
correlate with observed growth ratios $Y_{t}/Y_{t-1}.$\newline
\newline
Models 3 and 4 also have better predictive out-of-sample performance for
week 21 than M1 and M2. For example, the 95\% crI for $Y_{rep,T+1}$ under
model 4 is (194169, 214482) comfortably including the actual value of
210099. By contrast models M1 and M2 tend to underpredict the future value,
though the 97.5\% point crI for M2 just includes the true value. \newline
\newline
\textbf{Discussion}\newline
\newline
Infection time series with explosive phases are of current concern due to
repeated epidemic waves of the COVID\ outbreak. However, some infections may
be characterized as endemic, and much recent commentary on COVID opines that
it will eventually become an endemic disease. As noted by Katzorakis (2022),
"an endemic infection is one in which overall rates are static --- not
rising, not falling". Such infections may nevertheless have serious health
impacts, examples being malaria and HIV. \newline
\newline
Models suitable for relatively stable infectious diseases may not need to
include any mechanism for sudden fluctuations, so justifying stationarity
assumptions on autoregressive parameters. For an AR1 normal linear time
series this would imply a constraint that the absolute autoregressive
coefficient be under 1. Of course, formal definitions and properties of
stationarity series have generally considered normal linear models, whereas
the above analysis has considered modelling infection counts, albeit using
identity links. \newline
\newline
Stationarity in count time series (under a Poisson distribution), for an
identity link, and with AR1 dependence on previous counts, is considered by
Fokianos (2011), Fokianos and Tj\o stheim (2011) and Fokianos et al (2020).
Thus with $y_{t}\thicksim Po(\mu _{t}),$ $\mu _{t}=d+by_{t-1},$ stationarity
holds for $|b|<1.$ By contrast, as discussed by Fahrmeier and Tutz (2001,
page 244), AR1\ dependence on untransformed previous counts when a log-link
is adopted leads to explosive behaviour for any positive value of AR
coefficient. If a log-link is adopted, one is instead led to models, called
log-linear by Fokianos (2011), involving log-transformed previous counts,
such as

$log(\mu _{t})=d+b\log (y_{t-1}+1).$\newline
In fact, the log-linear representation can be adapted to space-time
infection data, but this has not been considered above. Another possible
extension in both the linear and log-linear cases includes a lag on the
intensity itself. Thus for the linear (identity link) case $\mu
_{t}=d+by_{t-1}+c\mu _{t-1},$ with again possible spatio-temporal
modification possible.\newline
\newline
For infection count time series with epidemic phases, stationarity is a
restrictive assumption, and allowing non-stationary values is appropriate
(e.g. Cazelles et al, 2018). The COVID epidemic has been characterised by
extended periods of relatively stable infection levels followed by sudden
sharp phases of increasing infection levels. Such epidemic peak are followed
by a period of descending rates and subsequent return to stability. The
above analysis has sought to allow an identity link autoregressive scheme to
adapt to these pronounced fluctuations, with departures from stationarity
during epidemic phases, but returning to stationarity as rates descend and
infection rates resume broad stability. In particular, the mixture link
model allows adaptivity to both explosive phases and static endemicity. 
\newline
\newline
The above analysis of two sets of area-time COVID\ infection data has shown
the mixture link approach, with a time varying weight that selects between
explosive or stationary options, provides a better fit to the observed data.
Both datasets include pronounced epidemic peaks as well as extended periods
with relatively stable infection levels. The link mixture approach has also
provided improved short term (one step ahead) predictions. Demand for such
forecasts has been a central feature of the COVID\ pandemic (Rosenfeld and
Tibshirani, 2021). \newline
\newline
The methodology proposed here may have application beyond infectious disease
counts, particularly to spatial and non-spatial panel data involving
considerable time fluctuations, and especially when positive temporal
autocorrelation or positive feedback from neighbouring locations is
anticipated on substantive grounds (Glaser, 2017). Possible examples include
urban crime (Liesenfeld et al, 2017), and spatial innovation diffusion
(Bivand, 2015).\newline
\newline
\textbf{References\newline
} \newline
Andrews, M., Tamura, K., Best, J, Ceasar, J, Batey, K, Kearse, T,
Powell-Wiley, T (2021). Spatial Clustering of County-Level COVID-19 Rates in
the US. International Journal of Environmental Research and Public Health,
18(22), 12170.\newline
Besag J., York J, Molli\'{e} A\ (1991) Bayesian image restoration with two
applications in spatial statistics. Ann. Inst. Statist. Math, 43(1): 1-59.%
\newline
Bivand, R. (2015) Spatial diffusion and spatial statistics: revisting H\"{a}%
gerstrand's study of innovation diffusion. Procedia Environmental Sciences,
27, 106-111.\newline
Brooks, S, Gelman, A. (1998) General methods for monitoring convergence of
iterative simulations. Journal of Computational and Graphical Statistics,
7(4): 434-455. \newline
Cazelles, B., Champagne, C, Dureau, J. (2018) Accounting for
non-stationarity in epidemiology by embedding time-varying parameters in
stochastic models. PLoS Computational Biology, 14(8), e1006211.\newline
Coly, S., Garrido, M., Abrial, D., Yao, A (2021) Bayesian hierarchical
models for disease mapping applied to contagious pathologies. PloS One,
16(1), e0222898.\newline
Czado C, Raftery A (2006) Choosing the link function and accounting for link
uncertainty in generalized linear models using Bayes factors. Statistical
Papers, 47, 419-442\newline
Czado, C., Gneiting, T, Held, L. (2009) Predictive model assessment for
count data. Biometrics, 65(4):1254--1261.\newline
Dalvi, A, Braga, J (2019) Spatial diffusion of the 2015--2016 Zika, dengue
and chikungunya epidemics in Rio de Janeiro Municipality, Brazil.
Epidemiology and Infection, 147: e237\newline
Dowdy, D, Golub, J, Chaisson, R, Saraceni, V. (2012) Heterogeneity in
tuberculosis transmission and the role of geographic hotspots in propagating
epidemics. Proceedings of the National Academy of Sciences, 109(24),
9557-9562.\newline
Fahrmeir L, Tutz G (2001) Multivariate Statistical Modelling Based on
Generalized Linear Models. Springer\newline
Fokianos, K (2011). Some recent progress in count time series. Statistics 45
(1): 49-58.\newline
Fokianos, K., Tj\o stheim, D. (2011). Log-linear Poisson autoregression.
Journal of Multivariate Analysis, 102(3), 563-578.\newline
Fokianos, K., St\o ve, B., Tj\o stheim, D., Doukhan, P. (2020). Multivariate
count autoregression. Bernoulli, 26(1), 471-499.\newline
Giuliani, D., Dickson, M, Espa, G., Santi, F (2020) Modelling and predicting
the spatio-temporal spread of COVID-19 in Italy. BMC Infectious Diseases,
20(1), 1-10.\newline
Glaser, S. (2017) A review of spatial econometric models for count data.
Hohenheim Discussion Papers in Business, Economics and Social Sciences, No.
19-2017\newline
Griffith, D, Li, B. (2021). Spatial-temporal modeling of initial COVID-19
diffusion: The cases of the Chinese Mainland and Conterminous United States.
Geo-spatial Information Science, 24(3), 340-362.\newline
Jalilian, A, Mateu, J (2021) A hierarchical spatio-temporal model to analyze
relative risk variations of COVID-19: a focus on Spain, Italy and Germany.
Stochastic Environmental Research and Risk Assessment, 35:797--812\newline
Katzourakis A\ (2022) COVID-19: endemic doesn't mean harmless. World View,
24 January 2022. Nature 601, 485 (2022)\newline
Lagazio, C., Dreassi, E., Biggeri, A. (2001) A hierarchical Bayesian model
for space-time variation of disease risk. Statistical Modelling, 1(1): 17-29.%
\newline
Lang, J (1999) Bayesian ordinal and binary regression models with a
parametric family of mixture links. Computational Statistics and Data
Analysis, 31: 59--87\newline
Lawson, A, Song, H (2010). Bayesian hierarchical modeling of the dynamics of
spatio-temporal influenza season outbreaks. Spatial and Spatio-temporal
Epidemiology, 1(2-3): 187-195.\newline
Liesenfeld R, Richard J, Vogler J (2017) Likelihood-Based Inference and
Prediction in Spatio-Temporal Panel Count Models for Urban Crimes. Journal
of Applied Econometrics, 32(3), 600-620.\newline
Lowe R, Lee S, O'Reilly K (2021) Combined effects of hydrometeorological
hazards and urbanisation on dengue risk in Brazil: a spatiotemporal
modelling study. Lancet Planet Health, 5: e209--19.\newline
Lunn D, Spiegelhalter D, Thomas A, Best N. (2009) The BUGS project:
Evolution, critique and future directions. Stat Med, 28(25):3049-67.\newline
Martines, M, Ferreira, R, Toppa, R, Assun\c{c}\~{a}o, L, Desjardins, M,
Delmelle, E (2021) Detecting space--time clusters of COVID-19 in Brazil:
mortality, inequality, socioeconomic vulnerability, and the relative risk of
the disease in Brazilian municipalities. Journal of Geographical Systems,
23(1): 7-36.\newline
Mitze, T., Kosfeld, R. (2021) The propagation effect of commuting to work in
the spatial transmission of COVID-19. Journal of Geographical Systems,
https://doi.org/10.1007/s10109-021-00349-3\newline
Office of National Statistics (2022) Coronavirus (COVID-19) Infection
Survey: England. Release Date 07 January 2022.
https://www.ons.gov.uk/peoplepopulationandcommunity/healthandsocialcare/conditionsanddiseases/datasets/coronaviruscovid19infectionsurveydata%
\newline
Paul M, Held L (2011) Predictive assessment of a non-linear random effects
model for multivariate time series of infectious disease counts. Statistics
in Medicine, 30(10), 1118-1136.\newline
Rosenfeld, R., Tibshirani, R. (2021). Epidemic tracking and forecasting:
lessons learned from a tumultuous year. Proceedings of the National Academy
of Sciences, 118(51)\newline
Rui R, Tian M, Tang, M, Ho, G, Wu, C (2021) Analysis of the spread of
COVID-19 in the USA with a spatio-temporal multivariate time series model.
International Journal of Environmental Research and Public Health, 18(2): 774%
\newline
Shand L., Li B, Park T, Albarrac\'{\i}n D (2018) Spatially varying
auto-regressive models for prediction of new human immunodeficiency virus
diagnoses. J Royal Stat Soc: Series C (Applied Statistics), 67(4): 1003-1022.%
\newline
Vahedi, B., Karimzadeh, M., Zoraghein, H. (2021) Spatiotemporal prediction
of COVID-19 cases using inter-and intra-county proxies of human
interactions. Nature Communications, 12(1), 1-15.\newline
Watanabe, S (2013) A Widely Applicable Bayesian Information Criterion.
Journal of Machine Learning Research. 14: 867--897\newline
Watson, S, Liu, Y., Lund, R, Gettings, J, Nordone, S, McMahan, C, Yabsley, M
(2017) A Bayesian spatio-temporal model for forecasting the prevalence of
antibodies to Borrelia burgdorferi, causative agent of Lyme disease, in
domestic dogs within the contiguous United States. PLoS One, 12(5), e0174428.%
\newline
\newline

\end{document}